# The Economic Costs of Containing a Pandemic

Asahi Noguchi[1]

## Abstract

The coronavirus disease (COVID-19) has caused one of the most serious social and economic losses to countries around the world since the Spanish influenza pandemic of 1918 (during World War I). It has resulted in enormous economic as well as social costs, such as increased deaths from the spread of infection in a region. This is because public regulations imposed by national and local governments to deter the spread of infection inevitably involves a deliberate suppression of the level of economic activity. Given this trade-off between economic activity and epidemic prevention, governments should execute public interventions to minimize social and economic losses from the pandemic. A major problem regarding the resultant economic losses is that it unequally impacts certain strata of the society. This raises an important question on how such economic losses should be shared equally across the society. At the same time, there is some antipathy towards economic compensation by means of public debt, which is likely to increase economic burden in the future. However, as Paul Samuelson once argued, much of the burden, whether due to public debt or otherwise, can only be borne by the present generation, and not by future generations.

## 1   Introduction

The coronavirus disease (COVID-19), which has spread to many countries around the world since the beginning of 2020, has inevitably caused great economic loss at a global scale, unlike any pandemic that has occurred since World War II. It is not simply due to the infectious nature of the disease. The weight of social losses created by the deaths of countless people due to the spread of the infection is enormous in modern society, with the exception of some dictatorships that still exist. On the basis of such modern value judgments, many countries are willing to implement measures such as lockdowns which clearly involve heavy economic losses—in order to save as many lives as possible.

---

[1] Senshu University, School of Economics, Tokyo, Japan

The Spanish influenza pandemic, which began in 1918 during World War I, resulted in cumulative deaths in the tens of millions throughout the world. In spite of considerable variation, even the smallest of the estimates was large enough to cause a temporary decline in the global population, which had previously been growing by around 13 million people per year (Roser [2020]). Even at that time, in some countries and regions, measures similar to the current lockdowns had been implemented. However, the immensely high death toll suggests that such conscious measures to prevent the spread of the disease may have been extremely rare. This might indicate not only the low level of medical knowledge, but also the low value attached to human life during that time.

In the COVID-19 pandemic, cities across numerous countries faced a rapid spread of infection, and almost without exception, severe restrictions on economic activity were implemented. This was because the only way to prevent further spread of infection was to reduce human contact as much as possible, which implied that economic activity should be kept to a minimum. These stark trade-offs between economic activity and epidemic prevention denote the general characteristics of policies for pandemic deterrence.

Another characteristic of modern policies for pandemic deterrence is that they often involve some form of economic compensation by governments and local authorities. This kind of public economic compensation basically has two policy objectives. One of these is the provision of financial incentives for leave of absence. If it were not for such financial incentives, people without sufficient savings would have no choice but to leave home and go to work to earn an income. The other is income compensation (similar to insurance) against the decline in corporate and household incomes that would inevitably result from policies to deter the spread of infection. This enables the economic costs of pandemic deterrence to be re-distributed more equitably across society.

Modern pandemic deterrence policies require huge public expenditures including public economic compensation. The relative size of this public spending could swell to levels not seen since the two world wars of the 20th century, in some cases as high as 30-40% of the GDP. In the case of COVID-19, most of the public spending associated with pandemic deterrence policies will probably also be financed by public debt. This is because governments do not want their economies, which have already contracted significantly due to pandemic deterrence policies, to atrophy further by raising taxes on people.

Nonetheless, concerns about ballooning public debt are sure to grow. There is a clear tendency among some national policy authorities to keep public economic compensation to a minimum, under the pretext of preventing the expansion of the burden in the future. There are many people who outrightly oppose public economic compensation because of the concern that its further expansion would directly increase the burden on themselves.

However, there is already a long history of debate and controversy in the economic science on the relationship between public debt and future burdens. While it was Abba Lerner who sparked the idea (Lerner [1948]), the more general insights were gleaned from his approach and later organized in a textbook by Paul Samuelson (Samuelson [1967]). Their argument conclusively demonstrates the economic truth that much of the economic burden, irrespective of whether the spending is financed by higher taxes or public debt, is ultimately borne by the current generation and not by the future generations.

This conclusion is essentially in harmony with respect to the public economic compensation that accompanies pandemic deterrence. In other words, even if the entire economic compensation is financed by public debt, it does not directly lead to a future burden. Economic compensation in the form of public debt is not intended to pass the burden on to future generations, but only to redistribute the burden among the current generation.

This paper proceeds as follows. Section 2 uses the concept of health capital to articulate the trade-offs between economic activity and epidemic prevention that societies should be aware of when trying to contain a pandemic. It also identifies the optimal public intervention path (in terms of minimizing the combined loss from pandemics as a weighted sum of social and economic losses) for national and local governments. Section 3 derives the economic losses caused by a pandemic response based on the trade-off relationship between economic activity and epidemic prevention, and subsequently identifies the need to equalize the unequal share of these economic losses across various strata of the society. Section 4 argues, mainly relying on Samuelson's argument, that the economic burden of pandemic preparedness is essentially borne by the current generation, and that even if the government's economic compensation was based on public debt, it would not be possible to transfer that burden to the future. Section 5 concludes that the various considerations made in relation to pandemic preparedness are both necessary and unnecessary, referring to the problems of balancing economic activity with epidemic prevention and sharing economic costs and the problems of increasing future burdens due to fiscal deterioration, respectively.

## 2   Economic analysis of counter-pandemic measures

### 2-1   Trade-off between economic activity and quarantine

As noted in Section 1, one distinct feature of pandemic preparedness is that it almost always has a trade-off relationship with economic activity. In peacetime when any threat of epidemic or pandemic is undetected, the policy idea of consciously suppressing economic activity to prevent the spread of infection does not exist. It is conceivable that there is some degree of trade-off between economic activity and disease prevention even in peacetime, although the society is not aware of it. However, once the epidemic or the pandemic becomes a reality, the situation changes drastically. The society is forced to realize that economic activity must be restricted more severely to strongly deter the spread of infection.

In the early stages of infection control, wherein a growing number of infections are being gradually reported overseas, but have not been confirmed inland, governments tend to adopt "bolder measures." These are aimed at strengthening the monitoring the influx of people from abroad, especially from countries and regions where the infection has been confirmed. Such measures would have a decidedly negative effect on domestic economic activity, as it would imply restricting foreign tourists and other people entering the country for business purposes. Nevertheless, the effect of bolder measures on domestic economic activity is likely to be very limited as long as they are successful. This is because there should be no need to strongly regulate economic activity in the country as long as the infection has not yet spread in the country.

However, as soon as it is confirmed that the infection has crossed the border into the country, there is a radical progression in phase. Then, the governments are forced to implement some kind of regulation. Typically, the governments provided instructions to the general public to refrain from activities, such as public gatherings and events; shut down various educational institutions, stores, and restaurants, where human contact is unavoidable; and transition to staggered attendance in companies.

Some countries and regions that were severely affected by the pandemic went beyond these directives or solely moral imperatives to implement certain legally enforceable public regulations, in the form of a lockdown, to deter the spread of the disease. Restrictions on mobility were imposed, except for the minimum activities necessary for the maintenance of life. Inevitably, all economic activities, other than the production and sale of goods and services that were considered to be basic necessities, were in principle suspended. In modern society, where the value of human life is highly respected, it is considered unavoidable to suffer a possible economic loss in order to save as many lives as possible, once faced with increasing human loss due to the spread of infection.

The relationship between economic activity and quarantine is illustrated in Figure 1. The vertical axis of this figure shows "income" or "production of goods and services" and the horizontal axis shows "health capital." In general, the income or output of goods and services at any given time is representative of the level of economic activity. In order to deter the spread of infection, it is necessary to ensure certain social environments, such as group avoidance through the promotion of absence from work (or remote work) and ensuring sufficient social distance.

Health capital comprises precisely these social environments.[2] There is clearly a trade-off between the two, since one must abandon some economic activity in order to secure them.

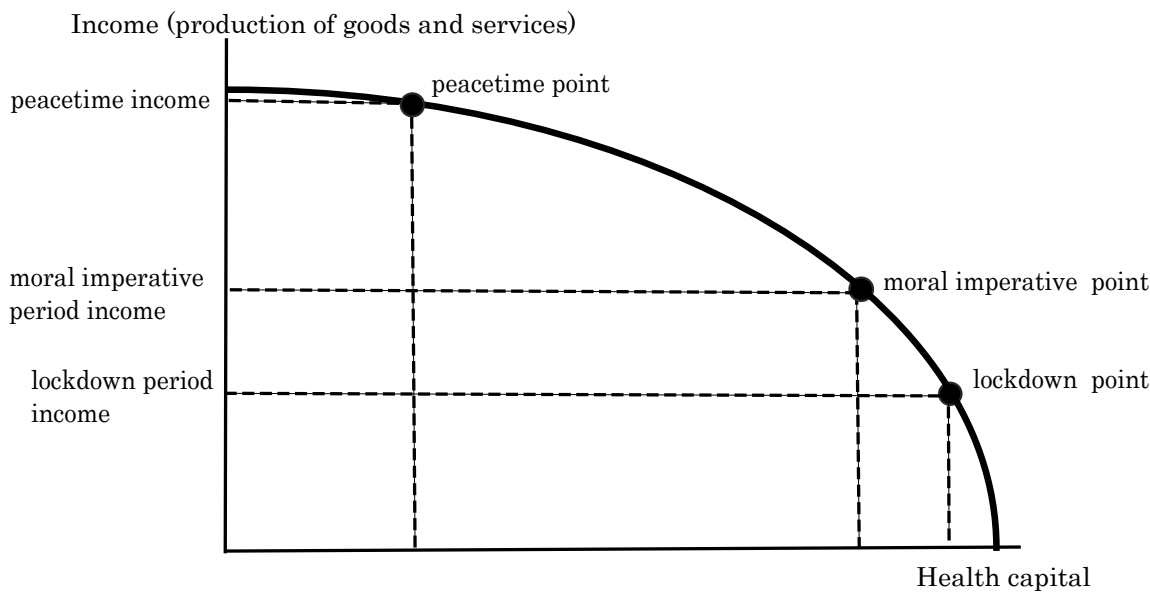

Figure 1  Trade-off between economic activity and quarantine

Three points are shown on this curve; peacetime, moral imperative, and lockdown. The trade-off between income and health capital does not arise significantly in the vicinity of the peacetime point. This implies that individuals can secure a certain amount of health capital without any significant income loss.

An ironic consequence of the COVID-19 infection control policies is that the imposition of lockdowns in large cities in developed countries (where economic activity had previously been very high) has resulted in significant improvements in the living conditions in the form of reduced air pollution. Some experts point out that these environmental improvements may have some effect of improving people's health. Conversely, this means that the vigorous economic activity that had been the norm prior to this pandemic had been worsening people's health through air pollution and other forms of environmental degradation, and the consequent social costs had been universally accepted as the price of economic benefit.

Such trade-offs between economic activity and health capital, although always existent, are rarely noticed by people in normal times. Most people do not feel any emergent need to secure health capital at the expense of apparent economic loss. However, securing health capital becomes a top priority for the society once a pandemic strikes. This is because the spread of infection in a pandemic would itself cause social losses, such as a rapid increase in the number of deaths due to the pandemic. Governments must then implement policies to curb all external activities of people, including economic activities, by means of public requests or legal regulations, in order to curb the spread of the infection.

---

[2] Grossman [1972] was the first to use the concept of health capital in economic analysis, in which health capital is grasped a kind of human capital embodied in the individual. Subsequently, however, there has been an emphasis on its public good nature in conjunction with the rise of the concept of "sustainability" in environmental and development economics, which is exemplified by Arrow, Dasgupta, and Mumford [2014].

In other words, the trade-offs between economic activity and health capital become more acute with the external environment, as the pandemic leads to a rapid expansion of socially needed health capital. In some cases, societies are forced to restrict all external activities other than the essential economic activities in order to ensure as much health capital as possible. This corresponds to the lockdown point in Figure 1. It means that society has made a public choice to indulge in the minimum income necessary to ensure the maximum health capital.

In contrast, the moral imperative point, is a situation in which the governments restrict business operations and individual mobility, not through legal coercion, but solely through some kind of social imperative. The declaration of a state of emergency issued by the Japanese government to major cities on April 7, 2020 (that was extended nationwide on April 16) is essentially such a measure. There is no legal regulation of people's behavior, such that health capital secured at that point is lesser than that at the lockdown point. There are also lesser constraints on economic activity and therefore, fewer economic losses. The Japanese government's decision to avoid the lockdown was likely due to its concerns about the resultant economic losses, as well as its constitutional restrictions. Similarly, Sweden decided to not implement lockdowns or mobility restrictions, but instead to limit to lenient regulations, such as requests for population bans, in order to acquire population immunity.

In the trade-off between maintaining economic activity and securing health capital for the protection of human life, there is always an implicit value judgment about how each society weights both the economic costs of deterring the spread of infection and the social costs of the spread. While the value of human life is extremely high in modern society, the economic cost of protecting that value cannot be neglected at all. That is the reason why even the most humanitarian nation will not ban cars, in spite of the fact that car accidents cause deaths around the world.

Some progressive intellectuals often argue that once the risk to human life becomes apparent, it should be addressed at any economic cost. Therefore, as regards COVID-19 disease control, all governments that do not impose lockdowns would be subject to criticism for their "disregard for human life." However, if even a small island nation with no confirmed cases of infection was forced to impose lockdown (unless risk of infection is absolutely zero), the islanders would be indignant at such interfering measure that is sure to disturb their daily living.

## 2-2 Social loss function of the pandemic

While the value judgement over the aforementioned trade-off is not necessarily a determinant, it always exists behind the policy decisions made by governments and local authorities. It is the very risk of infection spread, in an epidemiological sense, that determines how strongly governments should regulate the economy.

Strong public regulation is needed in the first place because the spread of the infection would be left unaddressed and the death toll would increase without it. Even if a government does not take any action against a pandemic because it is wary of economic loss, it is possible that the infection eventually subsides, and the number of deaths decreases as the population gains immunity. However, the cumulative death toll would be enormous, as it once was with the Spanish influenza. In contrast, if the government responds in some manner to a pandemic, it can certainly reduce the cumulative death toll, albeit with economic losses. However, the effect of the public intervention, i.e., the reduction in cumulative deaths, depends not only on the intensity of the intervention, such as whether it is requested or enforced, but also on the phase of the infection.

In other words, the nature and intensity of public interventions needed to deter the spread of infection clearly depends largely on the phase in which they are undertaken. This is evident by recalling the countries and regions, such as Taiwan and Hong Kong, that have had notable successes in their policies to prevent spread of COVID-19. Taiwan and Hong Kong have avoided serious outbreaks of COVID-19, such as those in the West, despite their deep ties to China, the original source of the outbreak. This is because their policymakers immediately grasped the seriousness of the situation from the information coming in from China and took measures, such as limiting entry at the border, from an early stage. As a result, Taiwan and Hong Kong, in contrast to Western countries, have avoided the implementation of measures, such as lockdowns, that result in massive economic losses. Since the infection had not spread to Taiwan and Hong Kong from the beginning, even if the lockdown had been implemented at this time, there would have been significant economic losses and little additional benefit from it.

Perhaps, the same can be said of the diminishing phase of the infection, after the peak of its spread has passed. The governments that have implemented lockdowns will endure the economic losses from restricting economic activity while they are doing it. However, if the additional social losses that would result from relaxing the regulations (i.e., the additional increase in cumulative deaths) were small enough, the governments would prefer to immediately relax the regulations and normalize economic activity. When this situation will occur is determined by health system measures, such as the establishment of effective countermeasures to prevent the spread of infection, the emergence of effective new drugs, and the acquisition of population immunity. If, as a result of such an improved epidemiological situation, the strength or weakness of public regulation to deter the spread of infection becomes almost indiscriminate against social losses, it would mean that the economy was once again in a situation where it should return to the peacetime point in Figure 1.

Therefore, the social losses caused by the spread of an epidemic can be expressed as follows:

$$SL = MSL - SG(P, I). \qquad (1)$$

This equation implies that the social loss caused by the pandemic $SL$ is the maximum social loss $MSL$ minus the social gain of the public intervention $SG$ that is a function of two variables: the phase of the infection at the time of the intervention $P$ and the intensity of the public intervention $I$. Hereafter, we will refer to this equation (1) as the social loss function of an epidemic or pandemic.

In general, social loss from the spread of an epidemic refers to all disadvantages other than the economic loss suffered due to the infection, the foremost of which is the loss of human life itself. If the social losses from a pandemic are represented by the cumulative number of deaths, then the value can be derived from the epidemic curve of the infection. This is because the cumulative number of deaths from a pandemic is merely the integral value of the newly infected, as illustrated in the epidemic curve, multiplied by the mortality rate.

Suppose that the epidemic curve of an infectious disease is mountain-shaped, as in a normal distribution. Moreover, suppose that there are four phases ($P=1$ to $P=4$) ranging from the beginning of the spread of the infection to its convergence. The intensity of public interventions to deter the spread of infection should also be represented by $I=0$ (no intervention), $I=1$ (weak intervention), and $I=2$ (strong intervention). Figure 2 shows the epidemic curve in the absence of any public intervention in all phases, and the phase categories from 1 to 4.

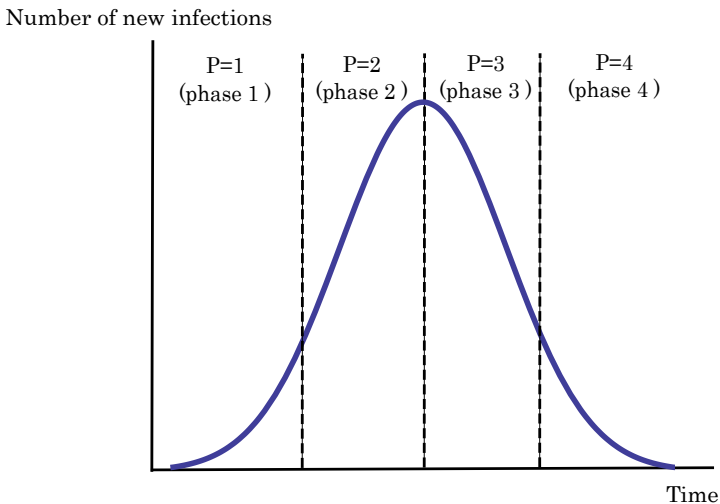

Figure 2  Epidemic curve and phases of the infection

The integral of this epidemic curve, i.e., the area of the mountain-shaped curve, represents the cumulative number of deaths in the absence of any public intervention in all the phases. That is the *MSL* in equation (1), which denotes the maximum social loss.

The shape of the epidemic curve will inevitably change if the government intervenes. However, the degree of change also depends on the phase of the infection. Figure 3 shows that, during the initial phase of the infection ($P=1$), the epidemic curve does not depend on the intensity of the public intervention. That is, if the intensity of the intervention in each phase is expressed as $I_P$, then the peak in $I_1=1$ is the same as that in $I_1=2$. Thus, even if Taiwan and Hong Kong were to implement lockdown in the early stages, the effect would not be much different from that of the border measures alone.

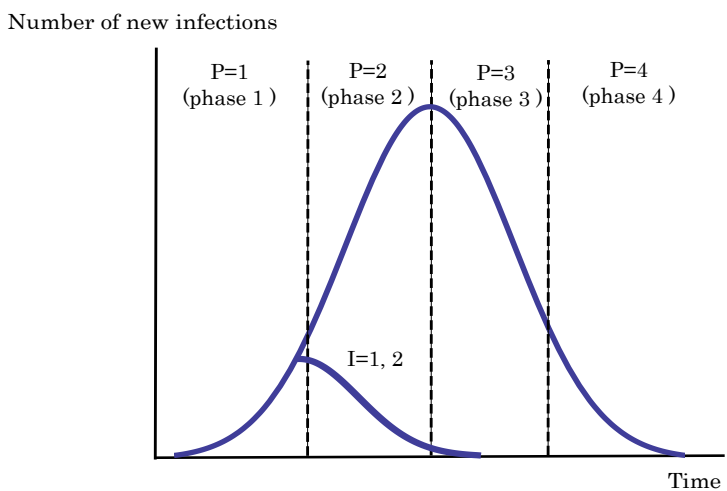

Figure 3  Epidemic curve in the case of an earlier intervention

However, the situation changes drastically at $P=2$, i.e., during the expanding phase of the infection. As Figure 4 shows, the shape of the epidemic curve is very different between a weak and a strong government intervention, i.e., between $I_2=1$ and $I_2=2$. Hence, the government needs to put in place stronger behavioral regulations, such as lockdowns, to curb the cumulative death toll from the infection.

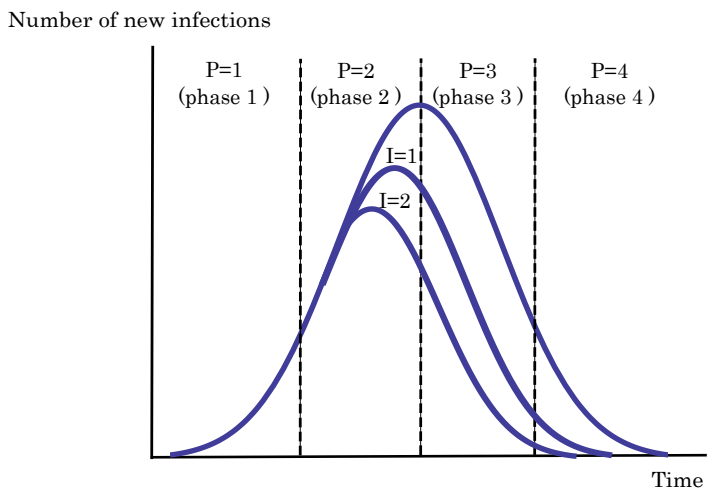

Figure 4  Epidemic curves in the case of weak and strong interventions

A similar dilemma could arise even during the phase where the spread of the infection has finally been deterred. Figure 5 depicts how the government's strong interventions in one stage led to a decline in new infections. If, however, the government does not bear the weight of the economic losses and weakens its intervention prematurely ($I_3$=1), then the spread of the infection may occur again. In such cases, it is likely that governments will be able to safely loosen the restrictions on people's behavior only after a phase, such as $P$=4, where the intensity of public intervention will no longer be a major determinant of the epidemic curve.

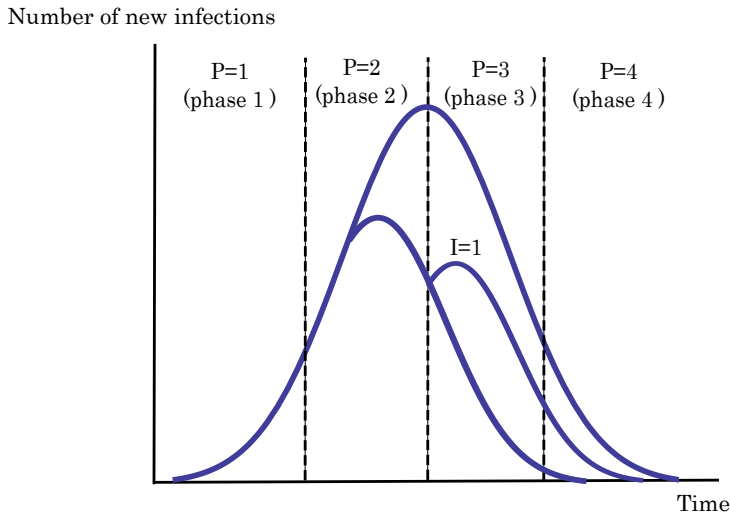

Figure 5  Epidemic curves in the case of premature weakening of interventions

From the above consideration, apart from the social loss $SL$ calculated at a point of public intervention with a certain intensity in a certain phase, the total social loss across all phases of infection $TSL$ can be obtained as follows:

$$TSL = MSL - \sum_{P=1}^{n} SG(P, I_P). \qquad (2)$$

## 2-3   Optimal public intervention path

As described above, governments generally choose the intensity of public interventions that is optimal for each phase, based on the assumption that the type of public interventions they undertake to deter the spread of infection will affect social losses represented by the cumulative number of deaths. Here, "optimal" does not necessarily mean minimizing social losses per se. In order to reduce social losses, more public intervention is needed to secure higher health capital, which entails greater economic loss in the form of further income loss. This is exactly the trade-off between economic activity and quarantine. What the government is trying to minimize is some combined measure of the economic and social loss. It can be expressed as follows:

$$CPL = EL - \lambda SL. \qquad (3)$$

This combined pandemic loss *CPL* is the weighted sum of the economic loss *EL* and social loss *SL* of this pandemic.[3] The coefficient λ is a parameter for converting the social loss (specifically, the loss of human life) to economic loss. The value represents each society's value judgment of human life. From equation (1), it is clear that the combined pandemic loss *CPL* is minimized at the point that the government determine the intensity of its intervention *I* so as to make the increment of economic loss (*ΔEL/ΔI*) equal to incremental value of social gain (λ*ΔSG/ΔI*), i.e., the value of the resulting reduction in the number of deaths.

At first glance, it may seem unethical to replace human life with an economic measure. However, our society has been constantly making certain implicit value judgments about the economic value of human life. The evidence of this is that economic activity is never halted, even though many lives are lost each year to asthma, which is due to the air pollution resulting from economic activity. Nevertheless, the economic value of human life is extremely high in many modern democratic societies; the degree to which this is the case is roughly proportional to the material wealth from economic growth. This is illustrated by the fact that many developed countries have suppressed economic activity to a minimum, albeit temporarily, to contain the pandemic.

If there is a dictatorship that does not recognize any value in the lives of its citizens, what choice will that government make in the event of a pandemic? In that case, the conversion rate λ for replacing human life with economic loss would be zero, and the total loss of the pandemic would be the economic loss itself. The best option for such a government would be to do nothing to deter the spread of the infection that would curtail economic activity, but to continue to let the infection continue as shown in Figure 2.

In ordinary democracies where the economic value of human life is high, choices about the nature and strength of public intervention are made after considering the economic losses resulting from the intervention in comparison to the social losses. However, as noted above, the choice depends on the phase on the epidemic curve where the public intervention takes place. Public interventions at an early stage on the epidemic curve, such as those in Taiwan and Hong Kong, can minimize pandemic losses without strong public interventions. The intervention path would be, for example, ($I_1$=1, $I_2$=1, $I_3$=0, $I_4$=0). It means that weak public interventions continue during the spread of the infection, but zero public interventions are used after it starts to shrink. The total combined loss resulting from the infection is the sum of the economic losses and the value of human losses borne at phase 1 and 2 only.

---

[3] This function is formally similar to the "central bank loss function" that is often used in macroeconomics to formulate the principles of central bank behavior. It is based on the idea that central banks typically conduct monetary policy in such a way so as to minimize social losses, defined by the weighted sum of inflation and the output gap. The trade-off between the inflation rate and the output gap, known as the Phillips curve, plays an important role.

In contrast, the optimal path when public intervention is used from the stage where the infection has already spread ($P=2$) would be, for example, ($I_1=0$, $I_2=2$, $I_3=1$, $I_4=0$). Here, the initial phase of the infection was left unattended, which forced strong public intervention in the spread phase. Even then, however, the degree of public intervention can be progressively relaxed as the infection lessens. This is because easing the behavioral restrictions did not increase the cumulative death toll much; the value of the additional human lives that would have been saved by strong restrictions fell short of the additional economic loss that would have been caused by them.

Thus, when public intervention is applied at a stage when the infection has already spread, the cumulative economic and human losses are large, and consequently, the total losses of a pandemic are very high. Even so, this is certainly the optimal choice, in the sense that it minimizes the total combined loss of the pandemic since the time of public intervention.

## 3    Economic losses from counter-pandemic measures and their sharing

As discussed in the previous section, governments minimize the combined value of the social and economic losses, by determining the path of public interventions for each phase of the epidemic curve, always taking into account the effect of their interventions on curbing the spread of infection and the extent to which they will result in economic losses. In this case, the social loss is the cumulative death toll from the pandemic; the economic loss is the difference between the average income in peacetime (before the pandemic occurred) and the lower income brought about by economic regulation due to the public intervention.

Figure 6 shows the magnitude of the economic costs incurred when the government locks down in the event of a pandemic. In a lockdown, any human activity that involves external contact is regulated, leaving only the minimum economic activity necessary to maintain people's livelihoods. Therefore, the income that is being brought in at that time can be considered the minimum required income in that economy. In other words, the economic cost of lockdown is the difference between the average income in peacetime and the minimum required income in the economy.[4] Realistically, if an economy growing at an average rate of 2% per year were to shift to zero growth due to the lockdown, the lost income (2%) would be considered the economic cost of the lockdown.

---

[4] However, this required minimum income may rise gradually over time. This is because people's livelihoods, which could be sustained in a short period of time by withdrawing their stocks of daily necessities, will become unsustainable if the lockdown is prolonged.

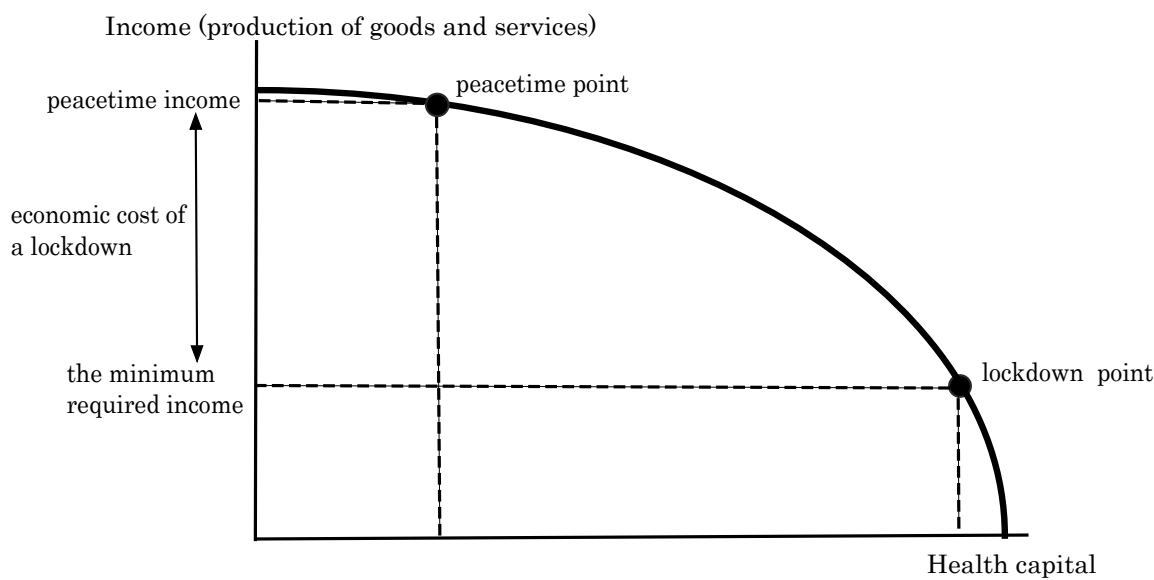

Figure 6  Economic cost of a lockdown

The path by which the decline in income and its growth rate come about will be dealt case-by-case. If the government halts business operations or firms' productive activities through legal regulation, the economic impact would appear to be a reduction in the supply of goods and services due to government regulation. In contrast, if the government asks people to avoid eating out, and people abstain from doing so in response, there should be a decrease in supply through the decrease in demand for food service. Thus, if the government merely calls for voluntary restraint, most of the income decline will be the result of demand shocks. However, the stronger the government regulation, the stronger will be the supply shock.

Importantly, all these consequences, whether they are demand shocks or supply shocks, unfold as a decline in people's current income, which constitutes the economic cost of pandemic preparedness, shown in Figure 6. In this sense, the question of whether the onset of decline is a demand or a supply shock is not so essential. The more pertinent question is about whose income it will reduce.

The government's counter-pandemic measures reduced the income of individuals and firms basically through reducing the supply of goods and services, whether directly or indirectly. In short, those who reduced the supply of goods and services faced income losses. In contrast, the individuals and firms that did not reduce the supply of goods and services did not have to contend with reduced incomes. This means that the economic costs of securing the health capital needed to prevent the spread of infection are being passed on in an extremely unequal manner among the members of society.

Thus, the government's counter-pandemic measures usually place a significant burden on some members of the society, in the form of loss of income, as the supply of goods and services shrinks. The distribution of that burden is extremely uneven with respect to industrial and employment conditions, because it depends on how government regulation affects firm and household incomes through demand and supply shocks. Hence, many governments that have dealt with pandemics have simultaneously offered some form of financial assistance, such as compensation for lost work or fixed benefits for companies and individuals who bear a particularly heavy burden.

While the policy objectives of such public support are not always clear, the following benefits are generally expected. The first is livelihood compensation for economically deprived households that have lost their income. This is primarily a measure aimed at low-income people, such as those with little to no savings. The second is the provision of economic incentives for leaves or absence from work. This becomes especially important when government measures remain a moral imperative rather than a legal regulation. In the absence of absenteeism or fixed benefits, there will always be people who are willing to go to work (instead of staying home) for financial reasons. The trend is particularly strong among low-income people who have no savings. In this sense, the government's support measures are not just focused on livelihood compensation, but also serve as an economic incentive to keep people at home, instead of permitting them to go out to work.

Another important role of the government's economic support is to enable the social sharing of the economic costs incurred due to the counter-pandemic measures. As noted above, the economic costs of preventing the spread of infection (or securing the necessary health capital for epidemic prevention) fall very unequally on certain members of the society. It is transferred to those who have reduced the supply of goods and services as a result of demand and supply shocks. They thereby lost the claims on goods and services from the present to the future.

However, those who have not reduced the supply of goods and services have also suffered economic welfare losses. This is because they have been forced to cut back on consumption due to the government's counter-pandemic measures, even though their incomes may not have reduced. Since they were inhibited to consume goods and services to the optimum level, their economic welfare has clearly declined. Since their incomes were not reduced, however, they necessarily increased their savings, which would result in their future claims on goods and services. In this regard, the economic welfare losses they have suffered might be almost negligible in comparison to the losses of those who have reduced their incomes.

In view of the above, the government's economic support should, in principle, be paid according to the share of the economic burden of fighting the pandemic (i.e., how much one's income is reduced by it). The government would be using its own funds to compensate for the damages based on an insurance-like principle, as argued by Hayashi [2020].

A major problem with this compensation policy is that it takes a long time to calculate the value of damages. If the government intervenes to combat the pandemic, many people will face an immediate reduction in their income. However, the manner in which the burden is ultimately passed on to the members of society can only be determined after the entire infection passes. For this reason, the government may be reluctant to provide financial support, which would result in low-income individuals with little to no savings pushed to further financial trouble.

It is not that difficult to solve these problems. Hayashi [2020] suggests that the government should provide unsecured bridge loans to households until the final compensation amount is finalized by the government. Alternatively, as many countries have already done, the funds can be distributed uniformly for the time being and adjusted afterwards through taxation. In that case, the government would make a distinction between those whose incomes have been significantly reduced by the government's counter-pandemic measures and those whose incomes have not been reduced, and subsequently give preferential tax treatment to the former.

## 4   Real and fictitious future burdens

### 4-1   Future burden from pandemic measures

The financial size of the public compensation could pile up to an amount not seen since wartime spending in World War II. Most of the public spending, at least for the time being, will be funded by public debt, and not by tax increases. This is because not many governments have the political capital to dare to raise taxes in the midst of the growing economic hardship.

Therefore, few mass media outlets and pundits have already begun to express concern over the future expansion of the budget deficit. Since the expansion of the budget deficit due to public support from the government is inevitable, such concerns will probably become more prevalent among the general public. This is because many people tend to assume that a larger government budget deficit means a larger future burden.

Government policies that deter the spread of infection may reduce not just people's current incomes, but also their future incomes. If the supply of capital goods is cut off due to the cessation of corporate production activities, capital investment by corporations and governments would become impossible, and capital formation in a country will stagnate. Moreover, if the accumulation and renewal of the capital stock is thus stalled, it will inevitably lead to stagnation of production and income in the future.

In addition, the expansion of leave of absence and furloughs that was implemented to deter the spread of infection will, if prolonged, lead to a degradation of human capital embodied in people's labor. Its probability becomes even stronger if dismissal of workers by companies increases. Furthermore, corporate bankruptcy due to a contraction of economic activity would imply a loss of management capital. Altogether, these shall result in a contraction of production and income in the future, rather than in the present.

However, these are all future costs that the society has no choice but to accept to deter the spread of the infection. They are never future burdens created by public support through government debt. On the contrary, if sufficient government support for businesses and households can effectively prevent the expansion of corporate bankruptcies and unemployment, the future burden created by the deterrence of the spread of the infection will certainly be reduced, at least to that extent.

## 4-2 The future burden of deficit financing

More generally, there are cases in which government support measures using deficit financing will lead to future burdens. It is a situation where deficit financing tightens capital markets, which in turn raises interest rates and crowds out private investment. There, to be sure, deficit financing, irrespective of how it is produced, usually stagnates capital formation and shrinks production and income in the future. However, in a situation where such a crowding-out mechanism does not arise, deficit financing, by itself, will not lead to a burden on future generations.

The economist who articulated this logic most clearly was Paul Samuelson, one of the leading economists of the 20th century, who discussed in detail the problem of future generations' burden of government deficit financing.[5] Samuelson first presents the conclusion of this issue as follows:

> The main ways that one generation can put a burden on a later generation is by using up currently the nation's stock of capital goods, or by failing to add the usual investment increment to the stock of capital. (Samuelson [1967] p.346)

The meaning of this brief statement is explained in more detail in a summary at the end of the chapter.

> The public debt does not burden the shoulders of a nation as if each citizen were made to carry rocks on his back. To the degree that we now follow policies of reduced capital formation which will pass on to posterity less capital goods, we can directly affect the production possibilities open to them. To the degree that we borrow from abroad for some transitory consumption purpose and pledge posterity to pay back the interest and principal on such external debt, we do place upon that posterity a net burden, which will be a subtraction from what they can later produce. To the degree that we bequeath to

---

[5] In the seventh edition of Samuelson's *Economics* published in 1967, the problem is mainly dealt in the second section of Chapter 19 entitled "The Public Debt and Modern Fiscal Policy," and in the appendix to Chapter 19 entitled "False and Genuine Burdens of the Public Debt."

posterity an internal debt but no change in capital stock beyond what would anyway have been given them, there may be various internal transfer effects as one group in the community receives a larger share of the goods then produced at the expense of another group. (Samuelson [1967] p.353)

This explanation by Samuelson can be generalized as the proposition that government debt is not a burden on future generations unless it reduces their future consumption possibilities. Conversely, government debt is a future generational burden only if it reduces the future consumption possibilities of people. This may be called the "Lerner-Samuelson proposition on the future generational burden of government deficit financing," since it was Abba Lerner who laid the foundation for this reasoning.

Lerner pointed out that whether public debt will be a burden for future generations depends on whether the national debt issued to finance the budget deficit will be absorbed domestically or abroad.[6] In the case of internal debt, even if future generations are taxed in order to redeem government bonds, they will still receive the payments, and their consumption possibilities as a whole will not necessarily decline.[7] In contrast, in the case of external debt, while the current generation does not have to cut spending, the future generations will have to truncate theirs (through government tax collection) to repay the debt to other citizens, and the consumption possibilities of future generations will necessarily be reduced by that amount.[8] Thus, external public debt will necessarily be a burden for future generations.

While Lerner's argument implies that future generational burden will necessarily arise as long as the public debt is external, it does not imply that it will not arise in principle in the case of internal debt. As Samuelson points out, even if all of the public debt is internal, the future possibilities of production and consumption will decrease if the renewal and accumulation of the capital stock is inhibited by the debt. Generally, deficit-financing policies undertaken in a full employment economy are more likely to result in tighter capital markets, higher interest rates, and a crowding out of private investment. Therefore, future generational

---

[6] It should be noted that "generation" in the discussion of Lerner and Samuelson means "all people living at a certain point in time", which is different from the usual generational concept of "all people born at a certain point in time."

[7] Lerner says of this, "Very few economists need to be reminded that if our children or grandchildren repay some of the national debt these payments will be made *to* our children or grandchildren and to nobody else. Taking them altogether they will no more be impoverished by making the repayments than they will be enriched by receiving them." (Lerner [1948] p.256)

[8] The distinction regarding whether the debt is internal or not should be made not by whether the bond is denominated in its own currency, but by whether it was purchased by a resident of the country who is subject to the government's power of tax collection.

burdens will certainly arise, even if the debts are internal.

The above consideration leads to a more precise version of the Lerner-Samuelson proposition that public debt will not be a burden on future generations if it is absorbed domestically and does not reduce the domestic capital stock, subsequently implying that there is no reduction in future consumption possibilities. This proposition holds strictly true in a Ricardian economy, wherein people behave as if they equate current tax increases with future tax increases due to bond issues. In this case, people's saving and spending behavior will not change whether government spending is financed by higher taxes or by bond issues. Therefore, more fiscal deficits mean neither higher interest rates nor greater external debts.

Even in non-Ricardian economies, the Lerner-Samuelson proposition holds, at least approximately, in an underemployment economy where unemployment due to lack of demand exists. As traditional Keynesian models (such as the IS-LM model) show, if the government implements a deficit fiscal policy, there will be some increase in interest rates and crowding out of private investment. However, the degree of crowding out will be contained because income and savings will increase due to the demand-led growth induced by fiscal spending.

In short, public debt does not necessarily imply a future generational burden. Whether or not the burden arises depends entirely on the economic situation. As Samuelson points out, the impact of the buildup of public debt as internal debt appears not so much as an expansion of the future burden itself, but rather as an income transfer between different strata of society.

Government bonds as public debt are nothing more than assets for bondholders and a right to acquire goods and services in the future. Thus, as long as government-issued bonds continue to be absorbed in the country, private sector assets will continue to grow. The economic conditions in the country will decide the macroeconomic effect of the expansion of private assets. In a Ricardian economy, this private sector asset signifies future tax increases, therefore would be completely offset by public debt. Thus, deficit financing would be macroeconomically neutral. This means that, as Robert Barro once argued, government debt cannot be the net wealth of the private sector (Barro [1974]). However, in the non-Ricardian economy on which traditional Keynesian economics has been premised, the economic expansionary effect of deficit-financing government spending is usually greater than tax-financing government spending.

The most obvious difference between the two arises in the domestic distribution of purchasing power. The holders of government bonds in the private sector should have relatively greater access to goods and services than those without them. This purchasing power effect does not depend on whether the economy is Ricardian or non- Ricardian, but on the decision of each economic agent about whether to cut their spending and buy the government-issued bonds that created the debt. It is merely the result of the intertemporal consumption choices of each economic agent—whether to cut current spending to obtain future goods and services or simply to desire current goods and services.

## 4-3    Inherent currentness in the burden of counter-pandemic measures

The important message in Samuelson's argument on the future burden of deficit financing is the economic truth that most of the burden created by government spending, whether it is financed by bond issues or tax increases, will ultimately be borne by the current generation, and not by the future generations. Samuelson uses the example of wartime costs to illustrate this. It is worth considering why this is so by following the context of the supplementary argument in Chapter 19 of *Economics* that "all debt came from World War II" (Samuelson [1967] p.356).

War usually requires artillery and ammunition, and the government spending for it is funded by either higher taxes or public debt. Suppose that all the costs were covered by bonds rather than taxes. Even then, the burden on future generations will not arise if all the public debt is absorbed domestically and does not result in the crowding out of private investment.

To confirm this, consider an economy in which there is no capital stock at all, and therefore no investment; people devote all their income from time to time to consumption. It is, so to speak, an economy that is based solely on hunting and gathering. In this consumption-only, investment-free economy, it is self-evident that no matter how much manpower is used in the production of artillery and ammunition, it will not result in a burden for future generations. While future generations could produce artillery and ammunition, they cannot bring it to the present unless they possess a time machine. Thus, in this case, the burden of producing artillery and ammunition would eventually be borne entirely by the current generation in the form of a reduction in its consumption. This is because there are always constraints on production resources, such as labor. As more human resources are allocated in the production of artillery and ammunition, the production of consumer goods must be reduced.

In conclusion, whether wartime costs are financed by higher taxes or public debt is merely a question of how that burden will be shared within the current generation. If it is financed by a tax increase, then the taxpayer who has reduced their disposable income will bear the cost through reduced consumption. In contrast, if it is financed by public bonds, the cost is borne by those who voluntarily reduce consumption and purchase the bonds.

However, this strong conclusion that all fiscal burdens are borne not by future generations but by the present generation is derived from the assumption that there is no foreign sector or capital stock; thus, it is by no means universally true. There will always be future burdens when public bonds are absorbed overseas. Even if they are absorbed internally, there will be future burdens if they reduce domestic investment and capital stock. Regardless of the burden on future generations, if there are individuals who must reduce their consumption due to the government's policy, it is clearly a burden for that generation of people. In this sense, the burden of the government's policy is considered to be borne by the current generation rather than future generations.

The same conclusion holds for the government's counter-pandemic measures. As shown in the previous section, imposing public regulations to prevent the spread of infection necessarily reduces the very level of economic activity, through a reduction in demand or supply. In a lockdown, the difference between the income in peacetime and in lockdown is precisely that economic loss (Figure 5). Thus, public regulations to prevent the spread of the infection affects the economy in the form of demand and supply shocks, shrinking the overall production of goods and services. However, the extent of the impact will vary depending on the circumstances in which each firm and household is located. While some areas will be devastated (such as the entertainment industry or the food and beverage industry), others will rather expand supply (such as the mask-making industry).

The economic burden thus created will be entirely borne by the present generation, whether or not it is financially supported by public debt (exactly as it is in the case of the burden of war expenditures). This is because it is the current generation of people who are reducing the production and consumption of goods and services through the imposed counter-pandemic measures. The amount of public debt that piles up at that time will not deter private investment as unemployment expands due to falling demand and interest rates continue to fall. Thus, future production and consumption possibilities are not depressed by the bond itself. Rather, these possibilities would be better than they would be without it if the government's economic support measures successfully prevented the expansion of corporate bankruptcies and unemployment. Apparently, these measures should be deficit-financed, and not involve tax, since they must be implemented without reducing people's disposable income.

In a wartime economy, people are sometimes forced to severely cut back on their consumption. In Japan, during World War II, moral slogans such as "luxury is the enemy" and "we don't want it, not until we win" were imposed on the people by the government. It comes from the economic principle that there is no free lunch; everything involves trade-offs. Therefore, to obtain the artillery and ammunition needed to carry out the war, the current generation has no choice but to reduce consumption and divert the productive resources (that were used to produce consumer goods) to the production of artillery and ammunition. The resulting decline in consumption is a burden on the current generation.

Meanwhile, in an economy under pandemic preparedness (such as lockdowns), the greatest burden for the current generation will come in the form of a decline in income. The present generation is subsequently involuntarily forced to produce non-income-generating leisure instead of producing artillery and ammunition. People would prefer to earn income by engaging in economic activities and enjoy consumption of goods and services from that income, rather than such excessive leisure. It is the current generation's income and consumption that is lost in this situation resulting in the economic burden of the counter-pandemic measures.

# 5      Relevance and irrelevance of various considerations regarding pandemic preparedness

The above considerations about the economic costs of pandemic deterrence, unfortunately, do not help us to derive some necessary policy measures. However, it does teach us what we should consider and what we should not be trapped in, when formulating a specific policy. This is important because, on this issue, there are things to consider and avoid at the same time.

When a society is faced with a pandemic, the preliminary decisions involve the measures to be used and the extent to which they will be imposed to deter the spread of the disease. Our goal is to minimize the losses that the society suffers. While the goal, in itself, is quite simple, the decision is far from simple. This is because the stronger the policy to prevent the spread of the disease, the greater the economic losses are likely to be. This is the trade-off between economic activity and epidemic prevention. It is precisely because of this trade-off that we continue to struggle with the decision to impose lockdowns once an infection has spread, and the duration for which it should continue. In fact, if the lockdown does not result in any economic loss, the society will happily continue to comply.

This trade-off between economic activity and epidemic prevention signifies that our decisions regarding pandemic deterrence are nothing less than judgments about our priorities. The extent to which a given quarantine policy is effective in deterring infection is a purely empirical epidemiological question that has nothing to do with value judgments. However, no matter how epidemiologically effective the policy may be, there is no immediate justification for its implementation if it involves an economic loss that results in income reduction. In order to implement the policy, a social consensus based on the exact value judgment is necessary; the benefits of reducing the social losses are sure to exceed the economic losses caused by it.

Another issue we must consider regarding the economic costs of pandemic deterrence is how it will ultimately be shared by the society. We see that the economic burden of policies to prevent the spread of infection falls on each class of the society in a very unequal way. It varies across industries, and it depends on an individual's employment category. Some people have lost their jobs and income entirely because of the policies; others have been able to maintain their existing income and enjoy their extra leisure time. Preventing the spread of infection is an obvious public good that benefits all members of the society, and yet, the cost is unequally shared. Therefore, governments should provide some form of financial support, such as absenteeism and fixed benefits, along with its policies to deter the spread of infection.

The biggest obstacle to such public support measures is a very universal concern that they will lead to a deterioration of government finances. The existence of such apprehensions may, in some cases, distort judgments about what and how far to go to deter the spread of the infection itself. It is plausible that some nations may lift the lockdown to avoid the expansion of public assistance, even though there remains a significant risk of the spread of the infection.

However, as Samuelson have argued, since many of the government's policies can only be carried out using current rather than future labor resources, much of the burden will eventually have to be borne by the current generation. Even if public bonds are issued to implement these measures, it does not mean that the burden will be passed on to the future generations. The extent of public debt merely changes the distribution of goods and services in the future according to the current saving and spending behavior of each economic agent.

This applies to the public assistance provided as part of the counter-pandemic measures. The current generation is already bearing the heavy burden of shrinking incomes and consumption. Even if all public support was provided by issuing public bonds, thereby expanding government debt, the burden of the current generation could not be passed on to future generations. Therefore, there is no basis for the notion that government economic support should be withheld as much as possible in order to avoid leaving a burden for future generations.

However, the above considerations about the future burden of deficit financing do not at all deny that the accumulation of public debt has any effect on the economy as a whole. Generally, if the economy were non-Ricardian, then an increase in public debt would act as a greater expansion of people's spending because it would mean an increase in private assets as a future claim to goods and services. Therefore, at some stage, governments and central banks may have to carry out macroeconomic tightening (using fiscal and monetary policies) to contain price increases due to the expansion of private spending. This suggests that the failure of governments and central banks to make such macroeconomic adjustments could lead to economic disruptions, such as fiscal collapse and soaring inflation.

Economists of the Keynesian position have been consistently emphasizing that government fiscal deficits are not generally evil, but rather that deficit financing during recessions is necessary for the stabilization of the macroeconomy. However, they did not deny the need for fiscal discipline in the sense of long-run fiscal equilibrium through the business cycle. This is because it is difficult even for Keynesians to deny, in principle, the possibility of macroeconomic turmoil that would occur if there were no fiscal discipline.[9]

Nevertheless, the claims made by some media outlets that the expansion of public support will lead to immediate fiscal collapse and hyperinflation can be considered a product of mere sensationalism. There is almost no chance that the excessive demand supported by deficit financing will make it difficult for governments and central banks to control inflation in a situation where many companies are facing a business downturn due to reduced consumer demand as a result of counter-pandemic measures. Rather, as Christopher Sims has suggested as a policy implication of the fiscal theory of price level (FTPL), the evils of being too captive to fiscal discipline are far greater in a demand-deficient, low interest rate economy than they would be otherwise (Sims [2016]).

---

[9] However, there is a position on the part of some Keynesians that denies even the need for fiscal equilibrium in the long-run sense. For example, modern money theory (MMT), proposed in mid-1990s by Warren Mosler, Randall Wray, and others, maintains that there is no inherent government fiscal constraint in an economy with a sovereign currency. They termed the rigid fiscal equilibrium doctrine as "deficit hawks," the cyclical deficit doctrine (on which the traditional Keynesians have relied) as "deficit doves," and their own position (that government finances do not even need to be balanced through the business cycle) as "deficit owls." (Mitchell, Wray, and Watts [2019] p.333-4)

There have been countless examples in history of countries falling into financial ruin due to excessive public debt (Reinhart and Rogoff [2009]). However, fiscal collapse has only occurred in very exceptional circumstances in developed countries with adequate tax collection capacity. Conversely, there are many examples that show noticeable "debt tolerance" in these countries. The United Kingdom had a public debt of 250% of GDP after the end of the World War II with no signs of fiscal collapse. Japan had experienced a sustained deterioration in public finances caused by the prolonged economic stagnation since the 1990s, which caused government bond rates to continue to fall, rather than rise.

In short, it is highly unlikely, at least in the developed world, that a transient increase in budget deficits due to an increase in public support will lead to immediate fiscal collapse or hyperinflation. Thus, in terms of pandemic preparedness, it is clearly not about what we should consider, but about what we should not.

# References

Arrow Kenneth J., Partha Dasgupta, and Kevin J. Mumford [2014] "Health Capital," in UNU-IHDP and UNEP, *Inclusive Wealth Report 2014: Measuring Progress toward Sustainability*, Cambridge: Cambridge University Press.

Barro, Robert J. [1974] "Are Government Bonds Net Wealth?" *Journal of Political Economy*, Vol.82, No.6, pp.1095-1117.

Hayashi, Fumio [2020], "Urgent Proposal for the Government Corona Emergency Economic Measures: Government Should Be Devoted to Life Insurance," (in Japanese) *Newsweek Japanese Edition*, April 8, 2020,
https://www.newsweekjapan.jp/stories/world/2020/04/post-93052.php

Grossman, Michael [1972[ "On the Concept of Health Capital and the Demand for Health," *Journal of Political Economy*, Vol.80, No.2, pp.223-255.

Lerner, Abba P. [1948] "The Burden of the National Debt," in Lloyd A. Metzler et al. (eds.), *Income, Employment and Public Policy, Essays in Honour of Alvin Hanson*, New York: W. W. Norton, pp.255-275.

Mitchell, William, L., Randall Wray, and Martin Watts [2019] *Macroeconomics*, London: Red Globe Press.

Reinhart, Carmen M. and Kenneth S. Rogoff [2009] *This Time Is Different: Eight Centuries of Financial Folly*, Princeton: Princeton University Press.

Roser, Max [2020] "The Spanish Flu (1918-20): The Global Impact of The Largest Influenza Pandemic in History," *Our World in Data*, https://ourworldindata.org/spanish-flu-largest-influenza-pandemic-in-history.

Samuelson, Paul A. [1967] *Economics: An Introductory Analysis*, 7th edition, New York: McGraw-Hill/Irwin.

Sims, Christopher [2016] "Fiscal Policy, Monetary Policy and Central Bank Independence,"
http://sims.princeton.edu/yftp/JacksonHole16/JHpaper.pdf.